\documentclass[aps,prd,twocolumn,superscriptaddress,showpacs,preprintnumbers,amsmath,amssymb]{revtex4}
\usepackage{graphicx}
\usepackage{dcolumn}
\usepackage{bm}
\graphicspath{{ps/}}

\begin{document}

\title{\boldmath 
Precise measurement of the $CP$ violation parameter
$\sin2\phi_1$ in $B^0\to(c\bar c)K^0$ decays
}

\affiliation{Budker Institute of Nuclear Physics SB RAS and Novosibirsk State University, Novosibirsk 630090}
\affiliation{Faculty of Mathematics and Physics, Charles University, Prague}
\affiliation{University of Cincinnati, Cincinnati, Ohio 45221}
\affiliation{Justus-Liebig-Universit\"at Gie\ss{}en, Gie\ss{}en}
\affiliation{Gifu University, Gifu}
\affiliation{Gyeongsang National University, Chinju}
\affiliation{Hanyang University, Seoul}
\affiliation{University of Hawaii, Honolulu, Hawaii 96822}
\affiliation{High Energy Accelerator Research Organization (KEK), Tsukuba}
\affiliation{Indian Institute of Technology Guwahati, Guwahati}
\affiliation{Institute of High Energy Physics, Chinese Academy of Sciences, Beijing}
\affiliation{Institute of High Energy Physics, Vienna}
\affiliation{Institute of High Energy Physics, Protvino}
\affiliation{Institute for Theoretical and Experimental Physics, Moscow}
\affiliation{J. Stefan Institute, Ljubljana}
\affiliation{Kanagawa University, Yokohama}
\affiliation{Institut f\"ur Experimentelle Kernphysik, Karlsruher Institut f\"ur Technologie, Karlsruhe}
\affiliation{Korea Institute of Science and Technology Information, Daejeon}
\affiliation{Korea University, Seoul}
\affiliation{Kyungpook National University, Taegu}
\affiliation{\'Ecole Polytechnique F\'ed\'erale de Lausanne (EPFL), Lausanne}
\affiliation{Faculty of Mathematics and Physics, University of Ljubljana, Ljubljana}
\affiliation{Luther College, Decorah, Iowa 52101}
\affiliation{University of Maribor, Maribor}
\affiliation{Max-Planck-Institut f\"ur Physik, M\"unchen}
\affiliation{University of Melbourne, School of Physics, Victoria 3010}
\affiliation{Graduate School of Science, Nagoya University, Nagoya}
\affiliation{Kobayashi-Maskawa Institute, Nagoya University, Nagoya}
\affiliation{Nara Women's University, Nara}
\affiliation{National United University, Miao Li}
\affiliation{Department of Physics, National Taiwan University, Taipei}
\affiliation{H. Niewodniczanski Institute of Nuclear Physics, Krakow}
\affiliation{Nippon Dental University, Niigata}
\affiliation{Niigata University, Niigata}
\affiliation{University of Nova Gorica, Nova Gorica}
\affiliation{Osaka City University, Osaka}
\affiliation{Pacific Northwest National Laboratory, Richland, Washington 99352}
\affiliation{Panjab University, Chandigarh}
\affiliation{Research Center for Nuclear Physics, Osaka University, Osaka}
\affiliation{University of Science and Technology of China, Hefei}
\affiliation{Seoul National University, Seoul}
\affiliation{Sungkyunkwan University, Suwon}
\affiliation{School of Physics, University of Sydney, NSW 2006}
\affiliation{Tata Institute of Fundamental Research, Mumbai}
\affiliation{Excellence Cluster Universe, Technische Universit\"at M\"unchen, Garching}
\affiliation{Toho University, Funabashi}
\affiliation{Tohoku Gakuin University, Tagajo}
\affiliation{Tohoku University, Sendai}
\affiliation{Department of Physics, University of Tokyo, Tokyo}
\affiliation{Tokyo Institute of Technology, Tokyo}
\affiliation{Tokyo Metropolitan University, Tokyo}
\affiliation{Tokyo University of Agriculture and Technology, Tokyo}
\affiliation{CNP, Virginia Polytechnic Institute and State University, Blacksburg, Virginia 24061}
\affiliation{Yamagata University, Yamagata}
\affiliation{Yonsei University, Seoul}
  \author{I.~Adachi}\affiliation{High Energy Accelerator Research Organization (KEK), Tsukuba} % KEK
  \author{H.~Aihara}\affiliation{Department of Physics, University of Tokyo, Tokyo} % Tokyo
  \author{D.~M.~Asner}\affiliation{Pacific Northwest National Laboratory, Richland, Washington 99352} % PNNL
  \author{V.~Aulchenko}\affiliation{Budker Institute of Nuclear Physics SB RAS and Novosibirsk State University, Novosibirsk 630090} % BINP
  \author{T.~Aushev}\affiliation{Institute for Theoretical and Experimental Physics, Moscow} % ITEP
  \author{T.~Aziz}\affiliation{Tata Institute of Fundamental Research, Mumbai} % Tata
  \author{A.~M.~Bakich}\affiliation{School of Physics, University of Sydney, NSW 2006} % Sydney
  \author{A.~Bay}\affiliation{\'Ecole Polytechnique F\'ed\'erale de Lausanne (EPFL), Lausanne} % Lausanne
  \author{V.~Bhardwaj}\affiliation{Nara Women's University, Nara} % Nara
  \author{B.~Bhuyan}\affiliation{Indian Institute of Technology Guwahati, Guwahati} % IITG
% ...
% BINP
\collaboration{The Belle Collaboration}

\begin{abstract}
We present a precise measurement of the $CP$ violation parameter
$\sin2\phi_1$ and the direct $CP$ violation parameter ${\cal A}_f$
using the final data sample of $772\times10^6$ $B\bar B$ pairs
collected at the $\Upsilon(4S)$ resonance with the Belle detector at
the KEKB asymmetric-energy $e^+e^-$ collider.  One neutral $B$ meson
is reconstructed in a $J/\psi K^0_S$, $\psi(2S)K^0_S$,
$\chi_{c1}K^0_S$ or $J/\psi K^0_L$ $CP$-eigenstate and its flavor is
identified from the decay products of the accompanying $B$ meson.
From the distribution of proper time intervals between the two $B$
decays, we obtain the following $CP$ violation parameters:
$\sin2\phi_1=0.667\pm0.023(\mbox{stat})\pm0.012(\mbox{syst})$ and
${\cal A}_f=0.006\pm0.016(\mbox{stat})\pm0.012(\mbox{syst})$.
\end{abstract}


\maketitle 

In the standard model (SM), $CP$ violation in the quark sector is
described by the Kobayashi-Maskawa (KM) theory~\cite{bib:km} in which
the quark-mixing matrix has a single irreducible complex phase that
gives rise to all $CP$-violating asymmetries.  In the decay chain
$\Upsilon(4S)\to B^0\bar B^0\to f_{CP}f_{\rm tag}$, where one of the
$B$ mesons decays at time $t_{CP}$ to a $CP$-eigenstate $f_{CP}$ and
the other $B$ meson decays at time $t_{\rm tag}$ to a final state $f_{\rm tag}$
that distinguishes between $B^0$ and $\bar B^0$, the decay rate has a
time dependence in the $\Upsilon(4S)$ rest frame~\cite{bib:sanda}
given by
\begin{eqnarray}
\label{eq:psig}
{\cal P}(\Delta t) = 
\frac{e^{-|\Delta t|/\tau_{B^0}}}{4\tau_{B^0}}
\biggl\{1 &+& q
\Bigl[ {\cal S}_f\sin(\Delta m_d\Delta t) \nonumber \\
   &+& {\cal A}_f\cos(\Delta m_d\Delta t)
\Bigr]
\biggr\}.
\end{eqnarray}
Here ${\cal S}_f$ and ${\cal A}_f$ are $CP$ violation parameters,
$\tau_{B^0}$ is the $B^0$ lifetime, $\Delta m_d$ is the mass
difference between the two neutral $B$ mass eigenstates, $\Delta
t\equiv t_{CP}-t_{\rm tag}$, and the $b$-flavor charge $q=+1~(-1)$
when the tagging $B$ meson is a $B^0$ ($\bar B^0$).  With very small
theoretical uncertainty~\cite{bib:sanda}, the SM predicts ${\cal
  S}_f=-\xi_f\sin2\phi_1$ and ${\cal A}_f=0$ for the $b\to c\bar cs$
transition, where $\xi_f=+1~(-1)$ corresponds to $CP$-even (-odd)
final states and $\phi_1$ is an interior angle of the KM unitarity
triangle, defined as
$\phi_1\equiv\mbox{arg}[-V_{cd}V^*_{cb}/V_{td}V^*_{tb}]$~\cite{bib:beta}.
The BaBar and Belle collaborations have published several
determinations of $\sin2\phi_1$ since the first 
observation~\cite{bib:babar_sin2beta_2001,bib:belle_sin2phi1_2001}; 
previous results used
$465\times10^6$~\cite{bib:babar_sin2beta_2009} and
$535\times10^6$~\cite{bib:belle_sin2phi1_2006} $B\bar B$ pairs,
respectively.

With recently available experimental results, not only $\sin2\phi_1$
but also other measurements of the sides of the unitarity triangle and
other $CP$ violation measurements make it possible to test the
consistency of the KM scheme.  The indirect determination of the angle
$\phi_1$ deviates by $2.7\sigma$ from the current world average for
the direct determination of $\sin2\phi_1$~\cite{bib:ckmfit}.
Equivalently, the $B^\pm\to\tau^\pm\nu_\tau$ branching fraction and
the resulting value of $|V_{ub}|$ differ by $2.8\sigma$ from the
prediction of the global fit~\cite{bib:ckmfit}, where the
$\sin2\phi_1$ value gives the most stringent constraint on the
indirect measurement.  Furthermore, time-dependent $CP$ violation in
the neutral $B$ meson decays mediated by flavor-changing $b\to s$
transitions may deviate from $CP$ violation in the $b\to c\bar cs$
case because of possible additional quantum
loops~\cite{bib:b2s_theory}.  To clarify whether new physics
contributes to $CP$-violating phenomena or $B^\pm\to\tau^\pm\nu_\tau$
decays, it is very important to determine $\sin2\phi_1$, the SM
reference, as precisely as possible.

In this Letter, we describe the final Belle measurement of
$\sin2\phi_1$ and ${\cal A}_f$ in $b\to c\bar cs$ induced $B$ decays
to $f_{CP}$.  The $B$ decays to the $CP$-odd eigenstates, $f_{CP}=$
$J/\psi K^0_S$, $\psi(2S)K^0_S$ and $\chi_{c1}K^0_S$, and the
$CP$-even eigenstate, $f_{CP}=J/\psi K^0_L$, are reconstructed using
$772\times10^6$ $B\bar B$ pairs, the entire data sample accumulated on
the $\Upsilon(4S)$ resonance with the Belle detector~\cite{bib:belle}
at the KEKB asymmetric-energy $e^+e^-$ collider~\cite{bib:kekb}.  Two
inner detector configurations were used. A 2.0 cm radius beampipe and
a 3-layer silicon vertex detector (SVD) were used for the first data
sample that contains $152\times10^6$ $B\bar B$ pairs.  The remaining
$620\times10^6$ $B\bar B$ pairs were accumulated with a 1.5 cm radius
beampipe, a 4-layer silicon vertex detector and a small-cell inner
drift chamber.  The latter data sample has been recently reprocessed
using a new charged track reconstruction algorithm, which
significantly increased the reconstruction efficiency for the
$B^0\to(c\bar c)K^0_S$ decay modes.  In particular, the gain for the
$B^0\to J/\psi K^0_S$ decay mode is 18\%.

The $\Upsilon(4S)$ is produced with a Lorentz boost of
$\beta\gamma=0.425$ nearly along the $z$-axis, which is antiparallel
to the positron beam direction.  Since the $B^0$ and $\bar B^0$ mesons
are approximately at rest in the $\Upsilon(4S)$ center-of-mass system
(CM), $\Delta t$ can be determined from the displacement in $z$
between the $f_{CP}$ and $f_{\rm tag}$ decay vertices: $\Delta
t\simeq(z_{CP}-z_{\rm tag})/(\beta\gamma c)\equiv\Delta z/(\beta\gamma
c)$.

Charged tracks reconstructed in the central drift chamber (CDC),
except for tracks from $K^0_S\to\pi^+\pi^-$ decays, are required to
originate from the interaction point (IP).  We distinguish charged
kaons from pions based on a kaon (pion) likelihood
$\mathcal{L}_{K(\pi)}$ derived from the time-of-flight scintillation
counters, aerogel threshold Cherenkov counters (ACC), and $dE/dx$
measurements in the CDC.  Electron identification is based on the
ratio of the electromagnetic calorimeter (ECL) cluster energy to the
particle momentum as well as a combination of $dE/dx$ measurements in
the CDC, the ACC response, and the position and shape of the
electromagnetic shower.  Muons are identified by track penetration
depth and hit scatter in the muon detector (KLM).  Photons are
identified as isolated ECL clusters that are not matched to any
charged track.

For the $J/\psi K^0_S$, $J/\psi K^0_L$ and $\psi(2S)K^0_S$ modes,
event selection is the same as in our previous
analyses~\cite{bib:belle_sin2phi1_2006,bib:belle_b2ccs_prev}, where
$J/\psi$ mesons are reconstructed via their decays to $\ell^+\ell^-$
($\ell=e,\mu$) and the $\psi(2S)$ mesons to $\ell^+\ell^-$ or
$J/\psi\pi^+\pi^-$.  For the modes $J/\psi K^0_L$ and
$\chi_{c1}K^0_S$, in which the $\chi_{c1}$ is reconstructed in the
$J/\psi\gamma$ final state, both $J/\psi$ daughter tracks must be
positively identified as leptons, whereas for the $J/\psi K^0_S$ and
$\psi(2S)K^0_S$ modes, at least one daughter must satisfy this
requirement.  Any other track having an ECL energy deposit consistent
with a minimum ionizing particle is accepted as a muon candidate and
any track satisfying either the $dE/dx$ or the ECL shower energy
requirements is retained as an electron candidate.  For $J/\psi\to
e^+e^-$ decays, the $e^\pm$ charmonium daughters are combined with
photons found within 50~mrad of the $e^+$ or $e^-$ direction in order
to account partially for final-state radiation and bremsstrahlung.  In
order to accommodate the remaining radiative tails, an asymmetric
invariant mass requirement is used to select $J/\psi$ and $\psi(2S)$
decays in dilepton modes,
$-150{\rm~MeV}/c^2<M_{e^+e^-}-M_{\psi}<36{\rm~MeV}/c^2$ and
$-60{\rm~MeV}/c^2<M_{\mu^+\mu^-}-M_{\psi}<36{\rm~MeV}/c^2$, where
$M_{\psi}$ denotes either the nominal $J/\psi$ or $\psi(2S)$ mass.
For $\psi(2S)\to J/\psi\pi^+\pi^-$ candidates, we require a mass
difference of
$580{\rm~MeV}/c^2<M_{\ell^+\ell^-\pi^+\pi^-}-M_{\ell^+\ell^-}<600{\rm~MeV}/c^2$,
and $\chi_{c1}\to J/\psi\gamma$ candidates are required to have a mass
difference of
$385.0{\rm~MeV}/c^2<M_{\ell^+\ell^-\gamma}-M_{\ell^+\ell^-}<430.5{\rm~MeV}/c^2$.
For each charmonium candidate, vertex-constrained and mass-constrained
fits are applied to improve its momentum resolution.

Candidate $K^0_S\to\pi^+\pi^-$ decays are selected by requirements on
their invariant mass, flight length and consistency between the
$K^0_S$ momentum direction and vertex position.  Candidate $K^0_L$
mesons are selected from ECL and/or KLM hit patterns that are
consistent with the presence of a shower induced by a $K^0_L$ meson.
The centroid of the $K^0_L$ candidate shower is required to be within
a $45^{\circ}$ cone centered on the $K^0_L$ direction calculated from
the two-body $B$ decay kinematics and the momentum of the
reconstructed $J/\psi$ meson.

For $B\to f_{CP}$ candidate reconstruction in modes other than $J/\psi
K^0_L$, $B$ candidates are identified by two kinematic variables: the
energy difference $\Delta E\equiv E_B^*-E_{\rm beam}^*$ and the
beam-energy constrained mass $M_{\rm bc}\equiv\sqrt{(E_{\rm
    beam}^*)^2-(p_B^*)^2}$, where $E_{\rm beam}^*$ is the CM beam
energy, and $E_B^*$ ($p_B^*$) is the CM energy (momentum) of the
reconstructed $B$ candidate.  The $B^0\to J/\psi K^0_L$ candidates are
identified by the value of $p_B^*$ calculated using a two-body decay
kinematic assumption.

The $b$-flavor of the accompanying $B$ meson is identified from
inclusive properties of particles that are not associated with the
reconstructed $B^0\to f_{CP}$ decay~\cite{bib:fbtg_nim}.  The tagging
information is represented by two parameters, the $b$-flavor charge
$q$ and purity $r$.  The parameter $r$ is an event-by-event,
MC-determined flavor-tagging dilution factor that ranges from $r=0$
for no flavor discrimination to $r=1$ for unambiguous flavor
assignment.  The data are sorted into seven intervals of $r$.  For
events with $r>0.1$, the wrong tag fractions for six $r$ intervals,
$w_l (l=1,6)$, and their differences between $B^0$ and $\bar B^0$
decays, $\Delta w_l$, are determined from semileptonic and hadronic
$b\to c$ decays~\cite{bib:belle_b2ccs_prev,bib:b2s_2005}.  If
$r\le0.1$, the wrong tag fraction is set to $0.5$, and therefore the
tagging information is not used.  The total effective tagging
efficiency, $\Sigma(f_l\times(1-2w_l)^2)$, is determined to be
$0.298\pm0.004$, where $f_l$ is the fraction of events in the category
$l$.

The vertex position for the $f_{CP}$ decay is reconstructed using
$J/\psi$ or $\psi(2S)$ daughter tracks that have a minimum number of
SVD hits, while the $f_{\rm tag}$ vertex is determined from
well-reconstructed tracks that are not assigned to
$f_{CP}$~\cite{bib:b2s_2005}.  A constraint on the IP profile in the
plane perpendicular to the $z$-axis is used with the selected tracks.
With this procedure, we are able to determine a vertex even in the
case where only one track has sufficient associated SVD hits.  The
fractions of the single track vertices for $f_{CP}$ and $f_{\rm tag}$
are about 12\% and 23\%, respectively.

For a single track vertex, the estimated error of the $z$ coordinate,
$\sigma_z$, is the indicator of the vertex fit quality and is required
to be less than $500~\mu$m.  On the other hand, a vertex reconstructed
using two or more tracks is characterized by a more robust
goodness-of-fit indicator.  In the previous
analysis~\cite{bib:belle_sin2phi1_2006}, the value of $\chi^2$ of the
vertex calculated solely along the $z$ direction was used. This is now
replaced by $h$, the value of $\chi^2$ in three-dimensional space
calculated using the charged tracks {\em without} using the
interaction-region profile's constraint~\cite{bib:prd_cck}.  A
detailed MC study indicates that $h$ is a superior indicator of the
vertex goodness-of-fit because it is less sensitive to the specific
$B$ decay mode; in particular, $h$ shows a smaller mode dependence for
the vertices reconstructed from $B\to J/\psi X$ and $B\to D^{(*)}X$
decays, which are used as control samples to determine the vertex
resolution parameters.  In the multiple-track vertex case, $h<50$ and
$\sigma_z<200~\mu$m are required.  For candidate events in which both
$B$ vertices are reconstructed, we retain only those events where the
$B$ vertices satisfy $|\Delta t|<70$~ps for further analysis.

\begin{figure}[htb]
\includegraphics[width=0.48\textwidth,clip]{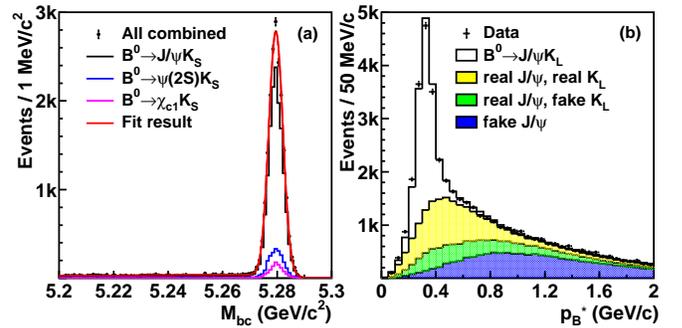}
\caption{(color online). (a) $M_{\rm bc}$ distribution within the
  $\Delta E$ signal region for $B^0\to J/\psi K^0_S$ (black),
  $\psi(2S)K^0_S$ (blue), and $\chi_{c1}K^0_S$ (magenta); the
  superimposed curve (red) shows the combined fit result for all these
  modes. (b) $p_B^*$ distribution of $B^0\to J/\psi K^0_L$ candidates
  with the results of the fit separately indicated as signal (open
  histogram), background with a real $J/\psi$ and real
  $K^0_L$'s (yellow), with a real $J/\psi$ and a
  fake $K^0_L$ candidate (green), and with a fake $J/\psi$ (blue).}
\label{fig:mbc_pb}
\end{figure}

For the candidate events in which both flavor tagging and vertex
reconstruction succeed, the signal yield and purity for each mode are
obtained from an unbinned maximum-likelihood fit to the
two-dimensional $\Delta E-M_{\rm bc}$ distribution for $f_{CP}$ modes
with a $K^0_S$ meson, and to the $p_B^*$ distribution for $J/\psi
K^0_L$.  The background mainly comes from $B\bar B$ events in which
one of the $B$ meson decays into a final state containing a correctly
reconstructed $J/\psi$, i.e., the $B\to J/\psi X$ process.  In order
to determine this background distribution, a $B\to J/\psi X$ MC sample
corresponding to 100 times the integrated luminosity of data is used.
An estimate of other combinatorial backgrounds is obtained from the
$M_{\ell^+\ell^-}$ sideband.  For $CP$-odd modes, the signal
distribution is modeled with a Gaussian function in $M_{\rm bc}$ and a
double Gaussian function in $\Delta E$.  The fits to determine signal
yields for these modes are performed in the region
$5.2{\rm~GeV}/c^2<M_{\rm bc}<5.3{\rm~GeV}/c^2$ and
$-0.1{\rm~GeV}<\Delta E<0.2{\rm~GeV}$.  The $p_B^*$ signal shape for
$J/\psi K^0_L$ is determined from MC events.  The requirement
$p_B^*<2.0{\rm~GeV}/c$ is used in the fit to estimate the signal yield
as well as the contribution of three categories of background: those
with a real (those that are correctly reconstructed) $J/\psi$ and a real
$K^0_L$, those with a real $J/\psi$ and a fake $K^0_L$ (those that 
are incorrectly reconstructed from electronic noise or electro-magnetic 
showers), and events with a fake $J/\psi$ (those that are background 
combinations).  The $M_{\rm bc}$ distribution for a stringent
$\Delta E$ requirement ($|\Delta E|<40$~MeV for $J/\psi K^0_S$,
$|\Delta E|<30$~MeV for $\psi(2S)K^0_S$ and $|\Delta E|<25$~MeV for
$\chi_{c1}K^0_S$) as well as the $p_B^*$ distribution for $J/\psi
K^0_L$ candidates are shown in Fig.~\ref{fig:mbc_pb}.  We require
$5.27{\rm~GeV}/c^2<M_{\rm bc}<5.29{\rm~GeV}/c^2$ for $f_{CP}$ modes
with a $K^0_S$ and $0.20{\rm~GeV}/c<p_B^*<0.45{\rm~GeV}/c$ for $J/\psi
K^0_L$ for the fit to the $CP$ violation parameters.  For the
candidates passing all the criteria mentioned above, the signal yield
and purity are estimated for each $CP$-eigenstate and listed in
Table~\ref{tab:sig_and_purity}.

\begin{table}[htb]
\caption{$CP$ eigenvalue ($\xi_f$), signal yield ($N_{\rm sig}$) and
  purity for each $B^0\to f_{CP}$ mode.}
\label{tab:sig_and_purity}
\begin{center}
\begin{tabular}{lccc}
\hline
\hline
Decay mode\hspace{1.8cm} & \,\,\,\,\,$\xi_f$\,\,\,\, & \,\,\,\,$N_{\rm sig}$\,\,\,\, & \,\,\,\,Purity (\%)\,\,\,\, \\
\hline
$J/\psi K^0_S$                     & $-1$ & 12649$\pm$114 & 97 \\
$\psi(2S)(\ell^+\ell^-) K^0_S$     & $-1$ &   904$\pm$ 31 & 92 \\
$\psi(2S)(J/\psi\pi^+\pi^-) K^0_S$ & $-1$ &  1067$\pm$ 33 & 90 \\
$\chi_{c1} K^0_S$                  & $-1$ &   940$\pm$ 33 & 86 \\
$J/\psi K^0_L$                     & $+1$ & 10040$\pm$154 & 63 \\
\hline
\hline
\end{tabular}
\end{center}
\end{table}

We determine ${\cal S}_f$ and ${\cal A}_f$ for each mode by performing
an unbinned maximum-likelihood fit to the observed $\Delta t$
distribution.  The probability density function (PDF) for the signal
distribution, ${\cal P}_{\rm sig}(\Delta t;{\cal S}_f,{\cal
  A}_f,q,w_l,\Delta w_l)$, is given by Eq.~(\ref{eq:psig}), fixing
$\tau_{B^0}$ and $\Delta m_d$ at their world average
values~\cite{bib:pdg2010} and including modifications to take the
effect of incorrect flavor assignment (parameterized by $w_l$ and
$\Delta w_l$) into account.  The distribution is convolved with the
proper-time interval resolution function, $R_{\rm sig}(\Delta t)$,
formed by convolving four components: the detector resolutions for
$z_{CP}$ and $z_{\rm tag}$, the shift of the $z_{\rm tag}$ vertex
position due to secondary tracks from charmed particle decays, and the
kinematic approximation that the $B$ mesons are at rest in the
CM frame~\cite{bib:resol_nim}.  Because we now use $h$ to
characterize the vertex goodness-of-fit, each of these resolution
function components in Ref.~\cite{bib:resol_nim} is reformulated as a
function of $h$ and $\sigma_z$.

Using the $M_{\rm bc}$ sideband events, the background PDF, ${\cal
  P}_{\rm bkg}(\Delta t)$, for each of the $CP$-odd modes is modeled
as a sum of exponential and prompt components, and is convolved with
$R_{\rm bkg}(\Delta t)$ expressed as a double Gaussian function.  In
the $J/\psi K^0_L$ mode, there are $CP$ violating modes among the
$B\to J/\psi X$ backgrounds, which are included in the background PDF.
The $\Delta t$ PDFs for the remaining $B\to J/\psi X$ and other
combinatorial backgrounds are estimated from the corresponding large
MC sample and $M_{\ell^+\ell^-}$ sideband events, respectively.  The
construction of these PDFs follows the same procedure as in our
previous analyses~\cite{bib:belle_sin2phi1_2006,bib:belle_b2ccs_prev}.

\begin{figure}[htb]
\includegraphics[width=0.237\textwidth,clip]{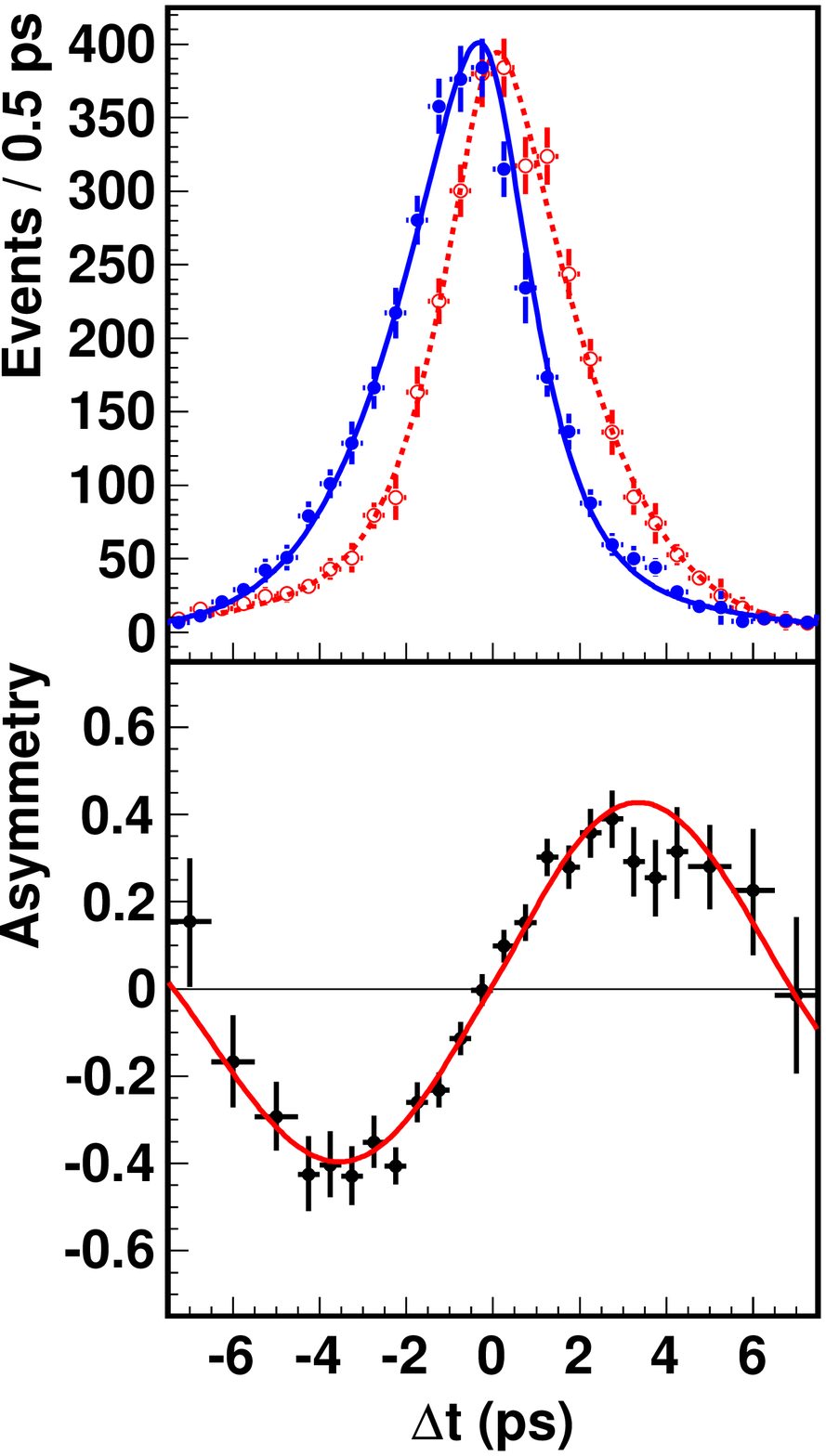}
\includegraphics[width=0.237\textwidth,clip]{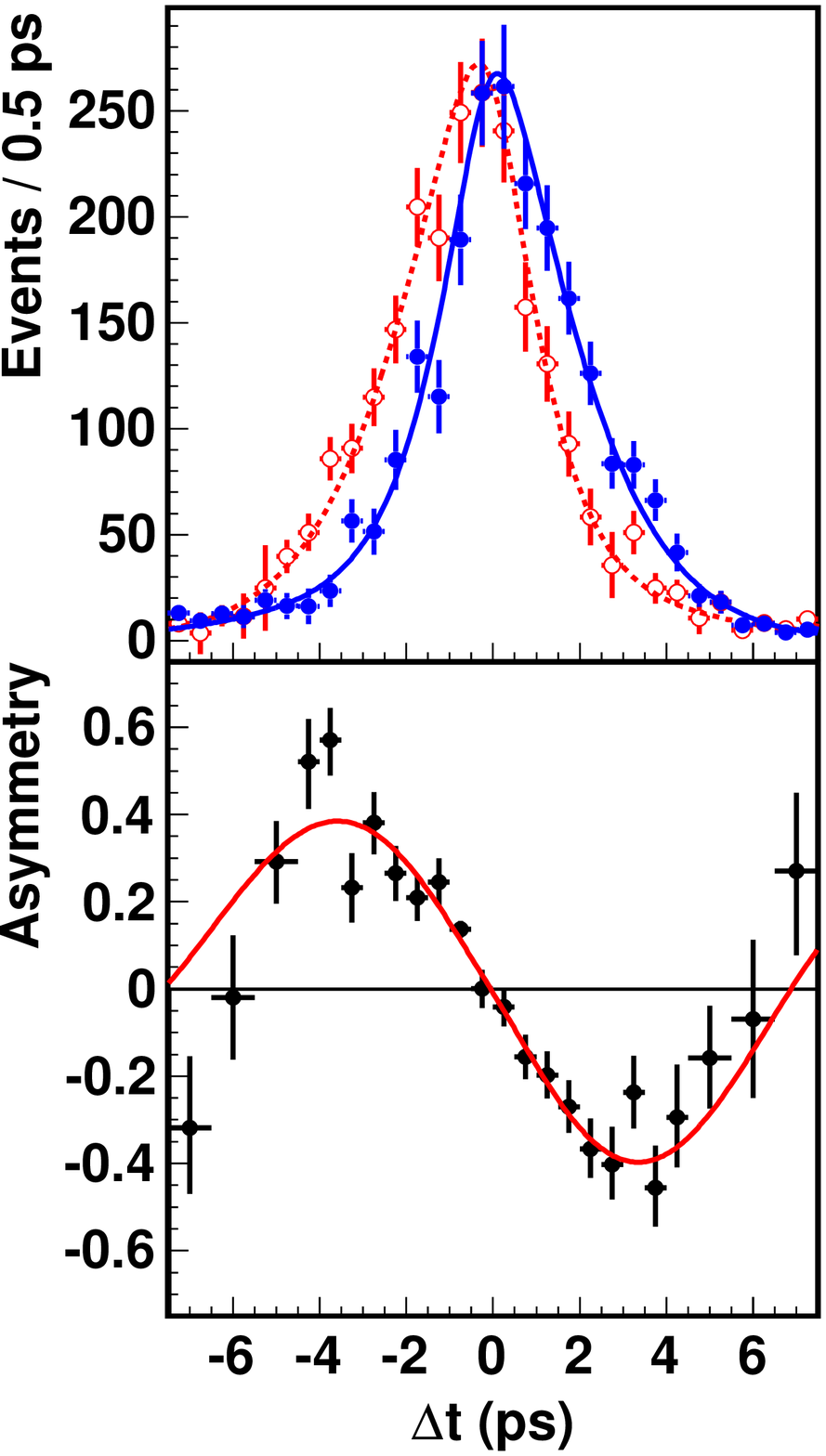}
\caption{(color online) The background-subtracted $\Delta t$
  distribution (top) for $q=+1$ (red) and $q=-1$ (blue) events and
  asymmetry (bottom) for good tag quality ($r>0.5$) events for all
  $CP$-odd modes combined (left) and the ${CP}$-even mode (right).}
\label{fig:dt_asymm}
\end{figure}

We determine the following likelihood for the $i$-th event:
\begin{eqnarray}
P_i = (1\!\!\!&-&\!\!\!f_{\rm ol})
\sum_kf_k \int\left[
{\cal P}_k(\Delta t')R_k(\Delta t_i-\Delta t')\right]d(\Delta t') \nonumber \\
&+&\!\!\!f_{\rm ol}P_{\rm ol}(\Delta t_i),
\label{eq:likelihood}
\end{eqnarray}
where the index $k$ labels each signal or background component.  The
fraction $f_k$ depends on the $r$ region and is calculated on an
event-by-event basis as a function of $\Delta E$ and $M_{\rm bc}$ for
the $CP$-odd modes and $p_B^*$ for the $CP$-even mode.  The term
$P_{\rm ol}(\Delta t)$ is a broad Gaussian function that represents an
outlier component $f_{\rm ol}$, which has a fractional normalization
of order 0.5\%~\cite{bib:resol_nim}.  The only free parameters in the
fits are ${\cal S}_f$ and ${\cal A}_f$, which are determined by
maximizing the likelihood function $L=\prod_iP_i(\Delta t_i;{\cal
  S}_f,{\cal A}_f)$.  This likelihood is maximized for each $f_{CP}$
mode individually, as well as for all modes combined taking into
account their $CP$-eigenstate values; the results are shown in
Table~\ref{tab:sin2phi1}.
Figure~\ref{fig:dt_asymm} shows the $\Delta t$
distributions and asymmetries for good tag quality ($r>0.5$) events.  
We define the background-subtracted asymmetry in each $\Delta t$ bin
by $(N_+-N_-)/(N_++N_-)$, where $N_+(N_-)$ is the signal yield with
$q=+1(-1)$.

\begin{table}[htb]
\caption{$CP$ violation parameters for each $B^0\to f_{CP}$ mode and
  from the simultaneous fit for all modes together.  The first and
  second errors are statistical and systematic uncertainties,
  respectively.}
\label{tab:sin2phi1}
\begin{center}
\begin{tabular}{lcc}
\hline
Decay mode        & $\sin2\phi_1\equiv-\xi_f{\cal S}_f$ & ${\cal A}_f$ \\
\hline
$J/\psi K^0_S$    & $+0.670\pm0.029\pm0.013$ & $-0.015\pm0.021^{+0.045}_{-0.023}$ \\
$\psi(2S) K^0_S$  & $+0.738\pm0.079\pm0.036$ & $+0.104\pm0.055^{+0.047}_{-0.027}$ \\
$\chi_{c1} K^0_S$ & $+0.640\pm0.117\pm0.040$ & $-0.017\pm0.083^{+0.046}_{-0.026}$ \\
$J/\psi K^0_L$    & $+0.642\pm0.047\pm0.021$ & $+0.019\pm0.026^{+0.017}_{-0.041}$ \\
\hline
All modes         & $+0.667\pm0.023\pm0.012$ & $+0.006 \pm 0.016 \pm 0.012$ \\
\hline
\end{tabular}
\end{center}
\end{table}

\begin{table}[htb]
\caption{Systematic errors in ${\cal S}_f$ and ${\cal A}_f$ in each
  $f_{CP}$ mode and for the sum of all modes.}
\label{tab:syst}
\begin{center}
\begin{tabular}{lcccccc}
\hline
             &  & $J/\psi K^0_S$ & $\psi(2S)K^0_S$ & $\chi_{c1}K^0_S$ & $J/\psi K^0_L$ & All \\
\hline
Vertexing    & ${\cal S}_f$ & $\pm0.008$ & $\pm0.031$ & $\pm0.025$ & $\pm0.011$ & $\pm0.007$ \\
             & ${\cal A}_f$ & $\pm0.022$ & $\pm0.026$ & $\pm0.021$ & $\pm0.015$ & $\pm0.007$ \\
\hline                                                                         
$\Delta t$   & ${\cal S}_f$ & $\pm0.007$ & $\pm0.007$ & $\pm0.005$ & $\pm0.007$ & $\pm0.007$ \\
resolution   & ${\cal A}_f$ & $\pm0.004$ & $\pm0.003$ & $\pm0.004$ & $\pm0.003$ & $\pm0.001$ \\
\hline                                                                         
Tag-side     & ${\cal S}_f$ & $\pm0.002$ & $\pm0.002$ & $\pm0.002$ & $\pm0.001$ & $\pm0.001$ \\
interference & ${\cal A}_f$ & $^{+0.038}_{-0.000}$ & $^{+0.038}_{-0.000}$ & $^{+0.038}_{-0.000}$ & $^{+0.000}_{-0.037}$ & $\pm0.008$ \\ 
\hline
Flavor       & ${\cal S}_f$ & $\pm0.003$ & $\pm0.003$ & $\pm0.004$ & $\pm0.003$ & $\pm0.004$ \\
tagging      & ${\cal A}_f$ & $\pm0.003$ & $\pm0.003$ & $\pm0.003$ & $\pm0.003$ & $\pm0.003$ \\
\hline                                                                         
Possible     & ${\cal S}_f$ & $\pm0.004$ & $\pm0.004$ & $\pm0.004$ & $\pm0.004$ & $\pm0.004$ \\
fit bias     & ${\cal A}_f$ & $\pm0.005$ & $\pm0.005$ & $\pm0.005$ & $\pm0.005$ & $\pm0.005$ \\
\hline                                                                         
Signal       & ${\cal S}_f$ & $\pm0.004$ & $\pm0.016$ &  $<0.001$  & $\pm0.016$ & $\pm0.004$ \\
fraction     & ${\cal A}_f$ & $\pm0.002$ & $\pm0.006$ &  $<0.001$  & $\pm0.006$ & $\pm0.002$ \\
\hline
Background      & ${\cal S}_f$ & $<0.001$ & $\pm0.002$ & $\pm0.030$ & $\pm0.002$ & $\pm0.001$ \\
$\Delta t$ PDFs & ${\cal A}_f$ & $<0.001$ &  $<0.001$  & $\pm0.014$ &  $<0.001$  &  $<0.001$  \\
\hline
Physics      & ${\cal S}_f$ & $\pm0.001$ & $\pm0.001$ & $\pm0.001$ & $\pm0.001$ & $\pm0.001$ \\
parameters   & ${\cal A}_f$ &  $<0.001$  &  $<0.001$  & $\pm0.001$ &  $<0.001$  &  $<0.001$  \\
\hline
Total        & ${\cal S}_f$ & $\pm0.013$ & $\pm0.036$ & $\pm0.040$ & $\pm0.021$ & $\pm0.012$ \\
             & ${\cal A}_f$ & $^{+0.045}_{-0.023}$ & $^{+0.047}_{-0.027}$ & $^{+0.046}_{-0.026}$ & $^{+0.017}_{-0.041}$ & $\pm0.012$ \\
\hline
\end{tabular}
\end{center}
\end{table}

Uncertainties originating from the vertex reconstruction algorithm are
a significant part of the systematic error for both $\sin2\phi_1$ and
${\cal A}_f$.  These uncertainties are reduced by almost a factor of
two compared to the previous analysis~\cite{bib:belle_sin2phi1_2006}
by using $h$ for the vertex-reconstruction goodness-of-fit parameter,
as described above.  In particular, the effect of the vertex quality
cut is estimated by changing the requirement to either $h<25$ or
$h<100$; the systematic error due to the IP constraint in the vertex
reconstruction is estimated by varying the IP profile size in the
plane perpendicular to the $z$-axis; the effect of the criterion for
the selection of tracks used in the $f_{\rm tag}$ vertex is estimated
by changing the requirement on the distance of closest approach with
respect to the reconstructed vertex by $\pm100~\mu$m from the nominal
maximum value of $500~\mu$m.  Systematic errors due to imperfect SVD
alignment are estimated from MC samples that have artificial
misalignment effects.  Small biases in the $\Delta z$ measurement are
observed in $e^+e^-\to\mu^+\mu^-$ and other control samples: to
account for these, a special correction function is applied and the
variation with respect to the nominal results is included as a
systematic error.  We also vary the $|\Delta t|$ range by $\pm30$ ps
to estimate the systematic uncertainty due to the $|\Delta t|$ fit
range.  The vertex resolution function is another major source of
$\sin2\phi_1$ and ${\cal A}_f$ uncertainty.  This effect is estimated
by varying each resolution function parameter obtained from data (MC)
by $\pm1\sigma$ ($\pm2\sigma$) and repeating the fit to add each
variation in quadrature.  The uncertainty in the estimated errors of
the parameters of reconstructed charged tracks is also taken into
account.  The largest contribution to the systematic uncertainty in
${\cal A}_f$ is the effect of the tag-side interference (TSI), which
is described in detail in~\cite{bib:tsi}.  Since the effect of TSI has
opposite sign for different $CP$-eigenstates, there is a partial
cancellation in the combined result.  Hence the combined TSI
systematic is smaller than the systematic in each individual mode.
Systematic errors due to uncertainties in the wrong-tag fractions are
studied by varying the wrong-tag fraction individually in each $r$
region.  A possible fit bias is examined by fitting a large number of
MC events.  Other contributions come from uncertainties in the signal
fractions, the background $\Delta t$ distribution, $\tau_{B^0}$ and
$\Delta m_d$.  Each contribution is summarized in
Table~\ref{tab:syst}.  We add them in quadrature to obtain the total
systematic uncertainty.

In summary, we present the final $\sin2\phi_1$ measurement using the
entire Belle $\Upsilon(4S)$ data sample containing
$772\times10^6~B\bar B$ pairs.  We have reconstructed $b\to c\bar cs$
induced $B$ meson decays in three $CP$-odd modes ($J/\psi K^0_S$,
$\psi(2S)K^0_S$, and $\chi_{c1}K^0_S$) and one $CP$-even mode ($J/\psi
K^0_L$).  The fit, using common $CP$-sensitive parameters for all four
modes, yields the values
$\sin2\phi_1=0.667\pm0.023(\mbox{stat})\pm0.012(\mbox{syst})$ and
${\cal A}_f=0.006\pm0.016(\mbox{stat})\pm0.012(\mbox{syst})$.
The results are consistent with previous 
measurements~\cite{bib:babar_sin2beta_2009,bib:belle_sin2phi1_2006}.  
These are the most precise determination of these parameters and solidify
the SM reference value used to test for evidence of new physics beyond
the SM.

We thank the KEKB group for excellent operation of the
accelerator; the KEK cryogenics group for efficient solenoid
operations; and the KEK computer group, the NII, and 
PNNL/EMSL for valuable computing and SINET4 network support.  
We acknowledge support from MEXT, JSPS and Nagoya's TLPRC (Japan);
ARC and DIISR (Australia); NSFC (China); MSMT (Czechia);
DST (India); INFN (Italy); MEST, NRF, GSDC of KISTI, and WCU (Korea); 
MNiSW (Poland); MES and RFAAE (Russia); ARRS (Slovenia); 
SNSF (Switzerland); NSC and MOE (Taiwan); and DOE and NSF (USA).


\begin{thebibliography}{999}
\bibitem{bib:beta}
Another naming convention $\beta(\equiv\phi_1)$ is also used in the literature.
\end{thebibliography}
\end{document}